\documentclass[aps,prc,superscriptaddress,showpacs,floatfix,nofootinbib,notitlepage,twocolumn]{revtex4-1}
\usepackage{amsmath,graphicx,float,latexsym,hyperref}
\def\bra{\langle}
\def\ket{\rangle}
\newcommand{\trento}{T$\mathrel{\protect\raisebox{-2.1pt}{R}}$ENTo}

\begin{document}
\title{Elliptic flow fluctuations in central collisions of spherical and deformed nuclei}

\author{Giuliano Giacalone}
\affiliation{Institut de physique th\'eorique, Universit\'e Paris-Saclay, CNRS,
CEA, F-91191 Gif-sur-Yvette, France}

\begin{abstract}
Elliptic flow ($v_2$) fluctuations in central heavy-ion collisions are direct probes of the fluctuating geometry of the quark-gluon plasma, and, as such, are strongly sensitive to any deviation from spherical symmetry in the shape of the colliding nuclei.
I investigate the consequences of nuclear deformation for $v_2$ fluctuations, and I assess whether current models of medium geometry are able to predict and capture such effects.
Assuming linear hydrodynamic response between $v_2$ and the eccentricity of the medium, $\varepsilon_2$, I perform accurate comparisons between model calculations of $\varepsilon_2$ fluctuations and STAR data on cumulants of elliptic flow, in central Au+Au and U+U collisions.
From these comparisons, I evince that the most distinct signatures of nuclear deformation appear in the non-Gaussianities of $v_2$ fluctuation, and I show, in particular, that the non-Gaussian $v_2$ fluctuations currently observed in central Au+Au collisions are incompatible with model calculations that implement a quadrupole coefficient of order 12\% in the $^{197}$Au nuclei.
Finally, I make robust predictions for the behavior of higher-order cumulants of $v_2$ in collisions of non-spherical nuclei.
\end{abstract}

\maketitle

\section{introduction}

Elliptic flow is the dynamical response of a fluid to an elliptic deformation of its geometry.
It is a salient feature of the hydrodynamic expansion of the quark-gluon plasma created in relativistic heavy-ion collisions, where elliptic anisotropy is generated, on the one hand, by the fact that the area of overlap of two nuclei at finite impact parameter looks like an ellipse~\cite{Ollitrault:1992bk}, which explains why elliptic flow grows quickly with the impact parameter of the collision, and on the other hand, by density fluctuations in the fluid~\cite{Alver:2008zza,Blaizot:2014nia}, that explain the striking emergence of elliptic flow in collisions at small impact parameter~\cite{Alver:2006wh}.
Following Teaney and Yan~\cite{Teaney:2010vd}, the elliptic anisotropy of the medium, dubbed $\varepsilon_2$, can be defined rigorously for a generic heavy-ion collision, and hydrodynamic simulations show that elliptic flow ($v_2$) is essentially a linear response to $\varepsilon_2$, i.e., $v_2=\kappa\varepsilon_2$, where $\kappa$ is a constant~\cite{Gardim:2011xv,Niemi:2012aj,Gardim:2014tya}.
This simple relation implies that $v_2$ and its event-by-event fluctuations can be used as direct probes of the fluctuating geometry of the quark-gluon plasma at the beginning of the hydrodynamic phase.

In central collisions, where linear hydrodynamic response is exhibited at its best~\cite{Niemi:2015qia,Noronha-Hostler:2015dbi}, an important source of fluctuations that contributes to $\varepsilon_2$ is given by the random orientation of the colliding nuclei, if they are non-spherical.
This explains the large magnitude of the rms elliptic flow measured in central collisions of nuclei that have a pronounced deformation, i.e., U+U collisions at the BNL Relativistic Heavy Ion Collider (RHIC)~\cite{Adamczyk:2015obl}, and Xe+Xe collisions at the CERN Large Hadron Collider (LHC)~\cite{Acharya:2018ihu}.
In the current modeling of initial conditions for hydrodynamics, the non-spherical shape of the nuclei is obtained by adding a quadrupole deformation in the wavefunctions of the colliding bodies, the implementation of which is performed with the guidance of tabulated data on nuclear ground-state deformations~\cite{Moller:1993ed,Moller:2015fba}.
This kind of modeling allows hydrodynamic simulations to reproduce quantitatively the aforementioned large rms elliptic flow observed in U+U~\cite{Goldschmidt:2015kpa} and Xe+Xe~\cite{Giacalone:2017dud,Eskola:2017bup} collisions.

In this paper, I argue that more interesting and nontrivial signatures of nuclear deformation can be observed in the non-Gaussian fluctuations of $v_2$, and that such effects are currently visible in the experimental data, in particular, in the fourth-order cumulant of elliptic flow $v_2\{4\}$, accurately measured by the STAR Collaboration~\cite{Adamczyk:2015obl} in central collisions of non-spherical nuclei, i.e., $^{197}$Au+$^{197}$Au collisions and $^{238}$U+$^{238}$U collisions. 
Note that the present study fills an important gap in the literature:
Theoretical studies devoted to the consequences of the prolate shape of $^{238}$U nuclei for relevant observables are numerous in the literature~\cite{Kolb:2000sd,Heinz:2004ir,Kuhlman:2005ts,Kuhlman:2006qp,Nepali:2006va,Masui:2009qk,Hirano:2010jg,Rybczynski:2012av,Schenke:2014tga,Shou:2014eya,Moreland:2014oya,Goldschmidt:2015kpa}, but a careful assessment of the impact of nuclear deformation on the non-Gaussianities of $v_2$ fluctuations, which are central to the phenomenology of flow in heavy-ion collisions~\cite{Aad:2014vba,Sirunyan:2017fts,Acharya:2018lmh,Adare:2018zkb}, is still missing.

I first investigate how nuclear deformation affects the fluctuations of eccentricity in Monte Carlo simulations, and whether these effects help explain the experimental data.
To achieve this, I use state-of-the-art Monte Carlo models of initial conditions (described in Sec.~\ref{sec:2}) to perform extensive calculations of $\varepsilon_2$ fluctuations, that I rescale and compare (Sec.~\ref{sec:3}) to STAR data on cumulants of elliptic flow fluctuations, $v_2\{2\}$ and $v_2\{4\}$, in central Au+Au and U+U collisions.
The outcome of these comparisons is eventually used (Sec.~\ref{sec:4}) to explain what are the prominent consequences of nuclear deformation for the non-Gaussian fluctuations of elliptic flow, and I predict how these effects can be observed in the higher-order cumulants of $v_2$.


\section{\label{sec:2} models of initial geometry}
In this section I aim at exhibiting models of initial conditions that are viable for hydrodynamic simulations of Au+Au and U+U collisions at~$\sqrt{s}=200$ GeV.
A model of initial conditions is a prescription that provides, event-by-event, the energy or entropy density, say $S(x,y)$, deposited in the transverse plane (for simplicity, at midrapidity) by a given collision.
This is by far the most crucial ingredient in the hydrodynamic framework, because it provides the medium with its spatial anisotropies\footnote{Note that for the definition of the eccentricity of the medium, $\varepsilon_2$, it does not matter whether one considers anisotropy in the energy density or in the entropy density of the fluid~\cite{Gardim:2011xv}.}, that get eventually converted into momentum anisotropies (i.e., the Fourier coefficients $v_n$) through the hydrodynamic evolution.

In this paper I shall employ two models of initial conditions, that correspond to two parametrizations of \trento{}~\cite{Moreland:2014oya}, a model for the profile of entropy density deposited at midrapidity in heavy-ion collisions.
Let me provide a detailed explanation of these models.

\subsection{\trento{} initial conditions}

\begin{figure*}[t!]
\centering
\includegraphics[width=\linewidth]{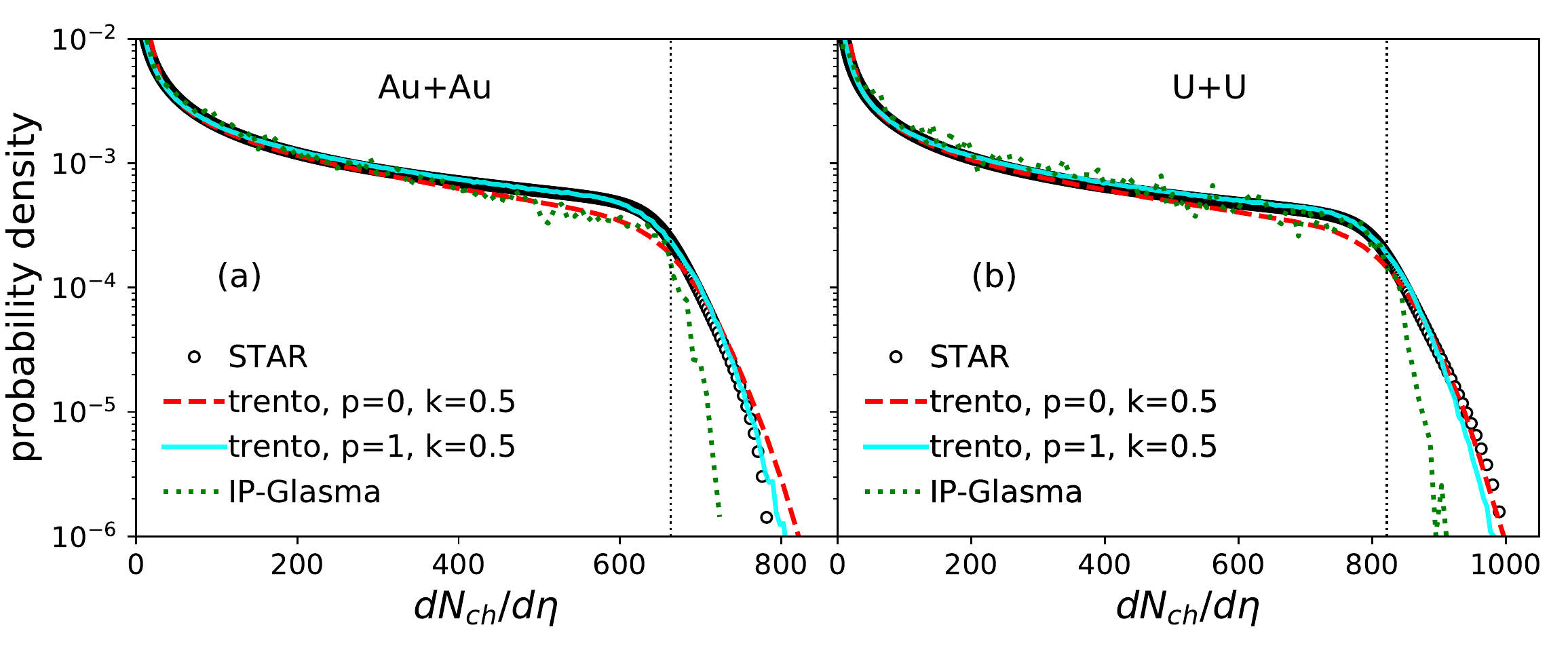}
\caption{Symbols: Distribution of multiplicity measured by the STAR collaboration \cite{Adamczyk:2015obl} in Au+Au collisions at ~$\sqrt[]{s}=200$ GeV [panel (a)], and U+U collisions at ~$\sqrt[]{s}=193$ GeV [panel (b)]. Lines are model calculations: IP-Glasma simulations~\cite{Schenke:2014tga} (dotted line), \trento{} with $p=1$ (solid line) and $p=0$ (dashed line). The vertical lines indicate the knees of the histograms.}
\label{fig:1}
\end{figure*}

The starting point is the modeling of the colliding bodies, which one performs through a random sampling of nucleons.
For a generic non-spherical nucleus, the sampling is done using the following Fermi distribution~\cite{Loizides:2014vua}:
\begin{equation}
\label{eq:ws}
\rho(x,y,z) = \frac{\rho_0}{1 + \exp \biggl[- \frac{1}{a} \biggl( r-R \bigl( 1+\beta_2 Y_{20} + \beta_4 Y_{40} \bigr) \biggr) \biggr]},
\end{equation}
where $r=\sqrt{x^2+y^2+z^2}$, $a$ and $R$ are the skin depth and radius of the nucleus, respectively, and the coefficients $\beta_2$ and $\beta_4$ multiplying the spherical harmonics, $Y_{20}=\sqrt[]{\frac{5}{16 \pi}}(3\cos^2 \theta-1)$, and $Y_{40}=\frac{3}{16 \sqrt{\pi}}(35\cos^4 \theta -30 \cos^2 \theta +3)$, are the coefficients of quadrupole and hexadecapole deformation of the nucleus, respectively.
Once the transverse coordinates of the nucleons, $(x,y)$, are sampled, the nuclei are overlapped at a random impact parameter, and nucleon-nucleon collisions take place.
One simply assumes that a participant nucleon deposits entropy, $s$, according to a Gaussian distribution:
\begin{equation}
\label{eq:gauss}
s_i (x,y)= \frac{\omega_i}{2\pi \sigma^2} \exp \biggl [ -\frac{(x-x_i)^2+(y-y_i)^2}{2\sigma^2} \biggr],
\end{equation}
where I take $\sigma=0.5$~fm for the width of each participant, and the index $i$ refers to the $i$-th participant nucleon.
The normalization, $\omega_i$, is randomly distributed according to the following gamma distribution:
\begin{equation}
P(\omega) = \frac{k^k \omega^{k-1}e^{-k}}{\Gamma(k)},
\end{equation}
which has mean value equal to unity, and variance proportional to $k^{-1}$.
Hence, the total entropy profile of a given nucleus, say A, is given by
\begin{equation}
S_A (x,y) = \sum_i s_i.
\end{equation}
Eventually, for two colliding nuclei, A and B, the total entropy profile of the system is given by a generalized mean, i.e.,
\begin{equation}
S(x,y)=\biggl ( \frac{S_A^p + S_B^p}{2} \biggr )^{1/p},
\end{equation}
where $p$ is any real number.

In this paper I shall use $S(x,y)$ from two different \trento{} parametrizations.
I take a geometric mean,
\begin{equation}
S=\biggl ( \frac{S_A^p + S_B^p}{2} \biggr )^{1/p}\biggl|_{p=0}=\sqrt[]{S_AS_B},
\end{equation}
and an arithmetic mean
\begin{equation}
S=\biggl ( \frac{S_A^p + S_B^p}{2} \biggr )^{1/p}\biggl|_{p=1}=\frac{S_A+S_B}{2}.
\end{equation}
Let me provide a bit of motivation for these choices.

The case $p=0$ is very successful in phenomenological applications, and is the favored value of $p$ resulting from the extensive Bayesian analyses of Refs.~\cite{Bernhard:2016tnd,Moreland:2018gsh}.
The reason of this success is likely the fact that $S$ in this model is proportional to the product $S_AS_B$.
This is reminiscent of a class of models inspired by high-energy QCD~\cite{Schenke:2012wb,Niemi:2015qia,Nagle:2018ybc}.
Taking $S(x,y)$ as a product leads typically to systems whose eccentricity follows closely the almond shape of the nuclear overlap~\cite{Yan:2014nsa}, leading eventually to a very good description of elliptic flow data~\cite{Moreland:2014oya,Giacalone:2017uqx}.

The case with $p=1$ corresponds instead to a wounded nucleon model with participant nucleon scaling~\cite{Rybczynski:2013yba}.
This class of models are variations of the original Monte Carlo Glauber~\cite{Loizides:2014vua} model, and are typically the models employed in experimental analyses\footnote{For instance, the model calculations shown by the STAR Collaboration in Ref.~\cite{Adamczyk:2015obl}, i.e., a Glauber Monte Carlo model with binary collision scaling, and a constituent quark Glaber Model model, are essentially mild variations of the $p=1$ model for what concerns the medium geometry, and should lead to the same kind of $\varepsilon_2$ fluctuations. This will be confirmed in Sec.~\ref{sec:3}.}.
Taking $S(x,y)$ as the sum of two components leads to systems that are more scattered in the transverse plane, and that have less enhanced eccentricity~\cite{Yan:2014nsa}.
The $p=1$ model is essentially ruled out by elliptic flow fluctuations data in Pb+Pb collisions at the LHC~\cite{Giacalone:2017uqx}.
Nevertheless, such a model has never been compared to RHIC data, and RHIC data was not used in the Bayesian analyses of Refs.~\cite{Bernhard:2016tnd,Moreland:2018gsh}.
I deem, then, that one can not \textit{a priori} state that the $p=1$ model is ruled out as well at RHIC energies.

The bottom line, and I shall come back on this point later, is that these two parametrizations of \trento{} capture, arguably, all the basic features of the widest classes of initial condition models for nucleus-nucleus collisions that are on the market.

\subsection{Multiplicity}
I discuss now the implementation of the parameter $k$ that regulates the fluctuations of the entropy produced by each participant nucleon.
This feature is important for a correct definition of collision centrality in the models.
In experiments, centrality classes are defined from the histogram of the multiplicity of particles (or hits, or energy measured in a calorimeter).
In hydrodynamics, one typically assumes that the total entropy at the initial condition is, in each event, proportional to multiplicity of particles in the final state.
Therefore, in numerical simulations one uses the histogram of total entropy to define the classes of collision centrality.
In order to exhibit meaningful model-to-data comparisons, I need to obtain in the models a distribution of total entropy which is in agreement with the distribution of $dN_{ch}/d\eta$ used by the STAR collaboration~\cite{Adamczyk:2015obl} to sort their events into centrality bins.
In particular, since I shall deal with central collisions, I want the high-multiplicity tails of the measured multiplicity distributions to be captured by the models.
This can be achieved by a proper choice of the fluctuation parameter, $k$.

In Fig.~\ref{fig:1}, I display as circles the distributions of $dN_{ch}/d\eta$ in Au+Au and U+U collisions measured by the STAR Collaboration\footnote{These distributions can be obtained from the parametrizations of $dN_{ch}/d\eta$ vs. centrality ($c$) provided at the beginning of Ref.~\cite{Adamczyk:2015obl}. Knowing $dN/d\eta(c)$, and using the fact that centrality is defined as the cumulative of the multiplicity, i.e., $c(dN/d\eta)=\int_{dN/d\eta}^{\infty}P(dN/d\eta)$, the plots of $P(dN/d\eta)$ shown in Fig.~\ref{fig:1} can be simply obtained by plotting $\biggl( \frac{d(dN/d\eta)}{dc} \biggr)^{-1}\biggl|_c$ vs. $dN/d\eta\bigl|_c$, with $c\in [0,1]$.}
Note that multiplicity distributions collected in the STAR detector present a high-multiplicity tail that is twice broader than that measured in detectors at the LHC~\cite{Das:2017ned}. 
This is consistent with the fact that detectors at RHIC have a much smaller acceptance, and therefore, they are more sensitive to statistical fluctuations.
Since in \trento{} the variance of fluctuations scales essentially as $1/k$, I expect that one needs to implement a lower value for the parameter $k$ at RHIC than at LHC.
The choice of this parameter is, essentially, detector-dependent.

Let me first discuss the case of the \trento{} model with $p=0$.
In Ref.~\cite{Giacalone:2017dud}, this model is found to provide an excellent description of ALICE data when $k=2.0$.
To describe RHIC data, i.e., a large-multiplicity tail which is twice broader, I use $k=0.5$, which is a factor 4 smaller, consistent with the fact that the width of the $\Gamma$ distribution is proportional to $k^{-1/2}$.
The probability distribution of the total entropy in this model is reported as a dashed line, for both Au+Au and U+U collisions, in Fig.~\ref{fig:1}.
To show a meaningful comparison between the distribution of entropy provided by \trento{} and the measured multiplicities, I rescale the entropy in \trento{} in order to have the \textit{knee} of the histogram at the same coordinate as the knee of the histogram of experimental data\footnote{The knee is defined as the mean value of multiplicity at zero impact parameter, and for the experimental data I calculate it through the fitting procedure of Ref.~\cite{Das:2017ned}. From the STAR parametrizations, I find $dN_{ch}/d\eta\bigl|_{\rm knee}=663$ in Au+Au collisions, and $dN_{ch}/d\eta\bigl|_{\rm knee}=821$ in U+U collisions.}.
Agreement with the data is reasonable\footnote{I have also checked that the model presents the same centrality of the knee, $c_{\rm knee}$, of data, i.e., the area of the histogram on the right of the knee. From the STAR parametrizations, I find $c_{\rm knee}=0.81\%$ in Au+Au collisions, and $c_{\rm knee}=0.75\%$ in U+U collisions.}, especially in the high-multiplicity tail, although the comparison is not as excellent as the one observed with LHC data\footnote{I have actually tried several values of $k$, and agreement does not get better.}.

Moving to \trento{} with $p=1$, I find that also in this case an excellent description of data is achieved using $k=0.5$.
The rescaled distributions of entropy in this model are shown as solid lines in Fig.~\ref{fig:1}.
Note that, for most of the histograms, the description provided by this model is better than that observed with $p=0$.

I would like to expand a bit further on this point, which is very striking.
The model with $p=0$ allows one to accurately reproduce LHC multiplicity data~\cite{Giacalone:2017dud}, but here it yields a poor description of RHIC data at multiplicities that are smaller than the location of the knee of the histogram.
At such multiplicities, one can not simply improve agreement with data via a change in $k$: The model is simply unable of reproducing the data.
Another model that is able to reproduce with great accuracy the distributions of multiplicity measured at LHC is the IP-Glasma model~\cite{Schenke:2012wb}, which was employed in calculations of Au+Au and U+U collisions in Ref.~\cite{Schenke:2014tga}.
Rescaled multiplicity distributions in this model are shown as dotted lines in Fig.~\ref{fig:1}.
Note that they are essentially compatible with the curves of \trento{} $p=0$, and consequently, this model provides a rather bad description of Au+Au data before the knee of the histogram\footnote{Note that the IP-Glasma results shown in Fig.~\ref{fig:1} present as well a large-multiplicity tail which is twice steeper than STAR data, suggesting that they miss the detector-dependent part of the fluctuations of multiplicity. In the IP-Glasma formalism, this issue could be solved via the inclusion of fluctuations of the saturation scale at the level of the colliding nucleons~\cite{McLerran:2015qxa}.}.
Therefore, either the difference between multiplicity distributions at RHIC and at LHC is due to some unknown issue due to the different detectors used, or there might actually be some physical motivation for which entropy production within IP-Glasma--like models (e.g. \trento{} $p=0$) provides a better description of data at the TeV energy scale.
Moreover, let me emphasize that the same issue might jeopardize future Bayesian analyses that, using the \trento{} model, are aimed at a determination of the parameters $k$ and $p$ via a simultaneous fit of RHIC and LHC data. 
My results here indicate that it is very unlikely that, using the same values of $k$ and $p$, one may be able to obtain a good description of both RHIC and LHC data.

That being said: 
I have exhibited two \trento{} parametrizations whose distributions of total entropy in Au+Au and U+U collisions provide a good description of the multiplicity distributions measured by the STAR Collaboration.
Therefore, these are viable models of initial conditions for hydrodynamic calculations, and I use them to compute the eccentricity of the medium in each event, as discussed below.

\subsection{Eccentricity fluctuations}
The elliptic anisotropy of a smooth profile in two dimensions, for instance, the entropy density $S(x,y)$ given in each collision by the \trento{} model, can be computed as indicated by Teaney and Yan~\cite{Teaney:2010vd}(in polar coordinates)\footnote{The expression for the anisotropy of order three (triangularity) is completely analogous~\cite{Teaney:2010vd}.}:
\begin{equation}
\label{eq:ecc2} \mathcal{E}_2 = \varepsilon_2e^{2i\Psi_2} =-\frac{\int r^2 e^{i 2 \phi }  S(r,\phi) r dr d\phi}{\int r^2 S(r,\phi) r dr d\phi}.
\end{equation}
This complex quantity fluctuates in both magnitude ($\varepsilon_2$) and orientation ($\Psi_2$) in each event.
Linear hydrodynamic response implies that $\mathcal{E}_2$ is linearly correlated with the complex elliptic flow coefficient, $V_2=v_2e^{2i\Phi_2}$, that also fluctuates in magnitude ($v_2$) and orientation ($\Phi_2$) event-to-event.
If this is the case, the probability distribution of $V_2$ coincides then with that of $\mathcal{E}_2$ up to a factor, and the statistical properties of $V_2$ fluctuations provide direct information about the fluctuations of the initial $\mathcal{E}_2$~\cite{Bhalerao:2011yg}.
Let me recall the formulas of the first two cumulants of the $v_2$ distribution:
\begin{align}
\nonumber v_2\{2\}&=\sqrt{\bra v_2^2 \ket},\\
\label{eq:kappa} v_2\{4\}&=\sqrt[4]{2\bra v_2^2 \ket^2 - \bra v_2^4 \ket},
\end{align}
where brackets indicate an average over events in a centrality bin.
Now, if $v_2=\kappa\varepsilon_2$, then,
\begin{align}
\nonumber v_2\{2\} = \kappa\varepsilon_2\{2\}, \\
v_2\{4\} = \kappa \varepsilon_2\{4\}.
\label{eq:cums}
\end{align}

The ratio $v_2\{4\}/v_2\{2\}$ has been used in many studies to observe, in hydrodynamic simulations, the transition between the linear regime, where $v_2\{4\}/v_2\{2\} = \varepsilon_2\{4\}/\varepsilon_2\{2\}$, and the nonlinear regime where this equality breaks down~\cite{Bhalerao:2011yg,Giacalone:2017uqx,Alba:2017hhe,Chattopadhyay:2017bjs}.
In Ref.~\cite{Giacalone:2017uqx}, in particular, the ratio $\varepsilon_2\{4\}/\varepsilon_2\{2\}$ was computed in extensive Monte Carlo calculations of $\varepsilon_2$ in very central collisions, and directly compared to LHC Pb+Pb data, in order to test the validity of different \trento{} parametrizations.

In what follows, I repeat this game, although without taking any ratio: 
I compute $\varepsilon_2\{2\}$ and $\varepsilon_2\{4\}$ in the \trento{} models, and then rescale them by an appropriate factor $\kappa$, in order to find the best possible agreement with experimental data on $v_2\{2\}$ and $v_2\{4\}$ in central Au+Au and U+U collisions.
As anticipated, eccentricity fluctuations in central collisions are strongly sensitive to the event-by-event fluctuations of the spatial orientation of the colliding $^{197}$Au and $^{238}$U nuclei, i.e., to the choice of the deformation parameters that enter in Eq.~(\ref{eq:ws}).
The goal of performing Monte Carlo calculations and model-to-data comparisons is essentially twofold.
First, I want to understand how $\varepsilon_2$ fluctuations are affected by the presence of deformed nuclei in the models. 
Second, I want to check whether these effects predicted by the models are in agreement with the features of $v_2$ fluctuations observed in the data.


\section{\label{sec:3} Comparison with STAR data}
\subsection{Setup}
Using the \trento{} parametrizations with $p=0$, $k=0.5$, and $p=1$, $k=0.5$, I simulate Au+Au and U+U collisions at~$\sqrt{s}=200$~GeV.
The nuclear shape parameters that enter in Eq.~(\ref{eq:ws}) are implemented as follows.
I take $^{238}$U nuclei with $R=6.80$~fm, $a=0.60$~fm, $\beta_2=0.236$, and $\beta_4=0.098$.
Concerning $^{197}$Au nuclei, I use $R = 6.40$ fm and $a=0.53$ fm, and I shall run calculations for both spherical nuclei, i.e., $\beta_2=\beta_4=0$, and deformed nuclei with $\beta_2=-0.125$, $\beta_4=-0.017$,

A couple of comments are in order.
The chosen values of radii and skin depths are rather standard.
I do not follow the suggestion of Ref.\cite{Shou:2014eya}, and do not set $a\sim0.4$ in both $^{238}$U and $^{197}$Au nuclei.
As suggested by the results shown in that same reference, this choice has essentially no impact on eccentricity fluctuations in central collisions.
Moving on to the deformation parameters, for $^{238}$U nuclei I implement a value of $\beta_2$ that is smaller than in previous studies, as I take it from the most recent table of nuclear deformations~\cite{Moller:2015fba}.
Moving to the deformation of $^{197}$Au nuclei, I cautiously stress that the parameter $\beta_2$ for this nucleus is not a measured quantity.
The data tables resulting from the model calculations of Refs.~\cite{Moller:1993ed,Moller:2015fba} yield $|\beta_2|\sim 0.13$, whereas, in Ref.~\cite{Nair:2008uw}, a value of 0.15 is guessed from the measured deformations of neighbor nuclei.
My choice $\beta_2=-0.125$ is taken again from Ref.~\cite{Moller:2015fba}.

For these \trento{} setups ($p=0$ and $p=1$), and using both spherical and deformed $^{197}$Au nuclei, I simulate {\cal O}($10^7)$ U+U and Au+Au collisions, and I compute the fluctuations of $\varepsilon_2$ as function of collision multiplicity, that I compare to experimental data.

When rescaling the results for $\varepsilon_2$ fluctuations, I shall take the same value of $\kappa$ [Eq.~(\ref{eq:cums})] for both Au+Au and U+U systems: 
This is a good approximation, as the value of $\kappa$ is reduced solely by viscous corrections.
Dimensional analysis indicates that viscous corrections scale like $1/R$, where $R$ is the radius of the system.
A good approximation for $R$ is given by $A^{-1/3}$.
Therefore, moving from U+U collisions to (smaller) Au+Au collisions, I expect a reduction of $\kappa$ of order $\sqrt[3]{238/197}$, which is a negligible 5\% correction.
Let me also stress that, in view of this argument concerning the system size, the value of $\kappa$ is expected to decrease as I move from central to peripheral collisions.
This effect should be very small in the centrality range I are interested in, as also indicated by hydrodynamic calculations~\cite{Noronha-Hostler:2015dbi}, therefore, I simply assume that $\kappa$ is a constant.

\subsection{Model vs. data}
I show STAR data~\cite{Adamczyk:2015obl} on cumulants of elliptic flow fluctuations as empty symbols in Fig.~\ref{fig:2}.
Figures~\ref{fig:2}(a) and~\ref{fig:2}(b) show STAR data in Au+Au collisions, whereas Figs.~\ref{fig:2}(c) and~\ref{fig:2}(d) show U+U collisions.
In each panel the minimum multiplicity corresponds to roughly 20\% centrality, whereas the maximum multiplicity is around 0.1\% centrality.

\subsubsection{Spherical $^{197}${\rm Au} nuclei}
\begin{figure*}[t!]
\centering
\includegraphics[width=.95\linewidth]{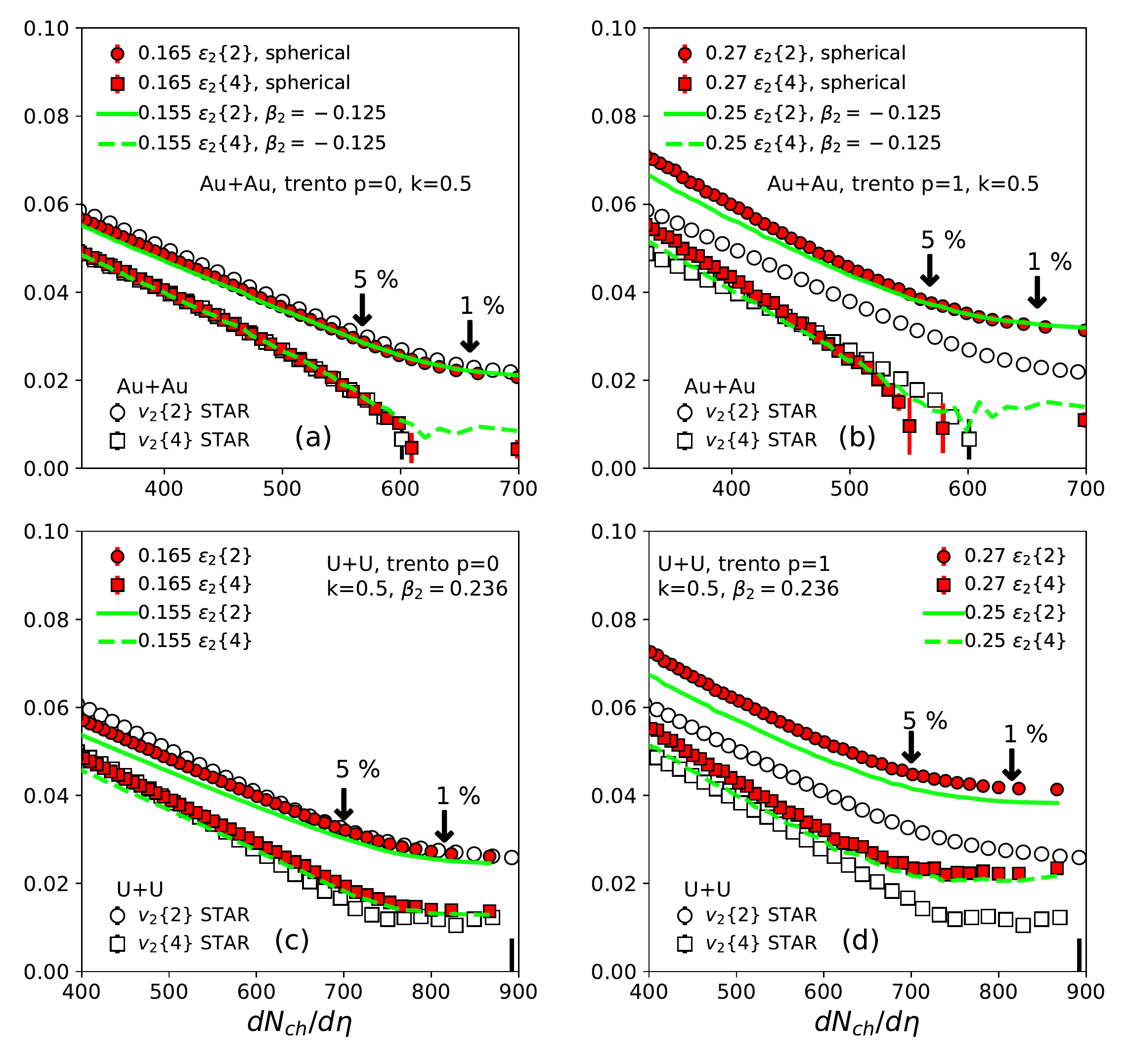}
\caption{Empty symbols: Cumulants of flow fluctuations measured by the STAR collaboration in central Au+Au collisions at $~\sqrt[]{s}=200$ GeV [panels (a) and (b)], and central U+U collisions at $~\sqrt[]{s}=193$ GeV [panel (c) and (d)], as function of collision multiplicity. Full symbols: Cumulants of initial eccentricity fluctuations in the \trento{} setup that implements spherical $^{197}$Au nuclei, with $p=0$ [panels (a) and (c)], and $p=1$ [panels (b) and (d)]. Lines: Cumulants of initial eccentricity fluctuations in the \trento{} setup that implements deformed $^{197}$Au nuclei, with $p=0$ [panels (a) and (c)], and $p=1$ [panels (b) and (d)]. The arrows indicate the correspondence between multiplicity and collision centrality.}
\label{fig:2}
\end{figure*}

I start by showing results from the \trento{} calculations implementing spherical $^{197}$Au nuclei.
These results are reported as full symbols in Fig.~\ref{fig:2}.

\begin{description}
\item[p=0]The calculations using \trento{} with $p=0$ are shown as red symbols in Figs.~\ref{fig:2}(a) and 2(c).
The value of the constant $\kappa=0.165$ is chosen in order to provide the most accurate possible description of the cumulant $v_2\{4\}$ in Au+Au collisions [panel (a)].
I choose to tune the model to $v_2\{4\}$ because 4-particle cumulants are not affected by nonflow contributions~\cite{Adare:2018zkb}.
The agreement between the model and the data in Au+Au collisions in Fig.~\ref{fig:2}(a) is excellent.
The description of $v_2\{4\}$ provided by this \trento{} parametrization is perfect.
It captures the trend of the data all the way up to 20\% centrality, and it correctly reproduces the change of sign of the cumulant $v_2\{4\}^4$ observed around 2.5\% centrality in experiment. 
Agreement is equally impressive for what concerns $v_2\{2\}$, all the way up to 20\%.
The slight shift of the experimental data towards larger values with respect to the model is easily understandable as due to nonflow.
Indeed, the non-flow subtraction performed by the STAR collaboration is not perfect, as two-particle correlations Ire calculated implementing a small gap of 0.1 units of pseudorapidity in the analysis~\cite{Adamczyk:2015obl}.
Moving to the red symbols shown in panel (c), the description of U+U data provided by this model is overall good, but less satisfactory.
I see that, although the comparison with data is not as quantitatively as good as in Au+Au collisions, the model captures nicely the qualitative features due to the deformation of $^{238}$U nuclei: 
$v_2\{2\}$ is larger in U+U than in Au+Au and, for reasons that I shall discuss in detail in Sec.~\ref{sec:4}, $v_2\{4\}$ is observed to be nonzero all the way to the highest multiplicities.

\item[p=1]I look now at the red symbols shown in Figs.~\ref{fig:2}(b) and~\ref{fig:2}(d), where I show results for \trento{} with $p=1$.
Again, the value of $\kappa=0.27$ is chosen such to yield the best description of $v_2\{4\}$ in Au+Au collisions\footnote{Note that the coefficient $\kappa$ varies a lot moving from $p=0$ to $p=1$. This is due to the fact that that the eccentricity grows much faster with centrality for $p=0$ than for $p=1$.}.
Starting with Au+Au collisions in Fig.~\ref{fig:2}(b), I see that the description of experimental data provided by this \trento{} parametrization is much worse than for the $p=0$ case:
$v_2\{2\}$ is overestimated, and $v_2\{4\}$ reaches zero at too large centrality with respect to the experimental data. 
This is consistent with the comparisons between this model and LHC data shown in Ref.~\cite{Giacalone:2017uqx}.
Agreement with data is also bad if I look at U+U collisions in Fig.~\ref{fig:2}(d), where a much smaller value of $\kappa$ would be needed to match the experimental data.
The qualitative features of U+U collisions are, on the other hand, reasonably captured, in particular, the fact that the cumulant $v_2\{4\}$ is nonzero all the way to the highest multiplicity.
\end{description}

\subsubsection{Deformed $^{197}${\rm Au} nuclei}
I discuss now the results obtained in \trento{} with deformed $^{197}$Au nuclei presenting $\beta_2=-0.125$ and $\beta_4=-0.017$~\cite{Moller:2015fba}.
These results are reported as lines in Fig.~\ref{fig:2}.

\begin{description} 
\item[p=0] The results for \trento{} with $p=0$ are the green lines in Figs.~\ref{fig:2}(a) and~\ref{fig:2}(c).  
The constant $\kappa$ that I need to tune the model to $v_2\{4\}$ in Au+Au collisions turns out to be smaller than previously with spherical $^{197}$Au nuclei, since both cumulants increase if I implement $\beta_2>0$.
Now, in Fig.~\ref{fig:2}(a) I see that the description of $v_2\{2\}$ is still very good all the way up to 20\% centrality, and so is the description of $v_2\{4\}$ essentially above 5\% centrality. 
But switching on the nuclear deformation has a dramatic effect of the fourth-order cumulant: 
It prevents $v_2\{4\}^4$ from going negative, and one observes a nonzero $v_2\{4\}$ all the way to the highest multiplicity, much as in the case of U+U collisions.
Nothing notable occurs for U+U collisions in Fig.~\ref{fig:2}(c).

\item[p=1]As for the results with $p=1$ with deformed $^{197}$Au nuclei, i.e., the lines shown in Figs.~\ref{fig:2}(b) and~\ref{fig:2}(d), I would simply like to point out is that I find the same striking result observed in Fig.~\ref{fig:2}(a): $v_2\{4\}$ is always nonzero in Au+Au collisions.
\end{description}

\subsection{Discussion}
Let me draw my conclusions from the comparisons shown in Fig.~\ref{fig:2}.

First, I confirm that the $p=1$ model is ruled out by elliptic flow fluctuations data at RHIC.
This is not a surprise:
First, this model is currently ruled out by LHC data~\cite{Giacalone:2017uqx}; Second, as anticipated, this model is similar to the models used by the STAR Collaboration in their analysis, and those models provide a bad description of data, in the sense that they present a $\kappa$ that varies with centrality even at the highest multiplicities.

Second, my results imply that a value $\beta_2\sim-0.12$ in $^{197}$Au nuclei is essentially ruled out by experimental data. 
The description of data achieved with the \trento{} $p=0$ parametrization implementing spherical $^{197}$Au nuclei is remarkably good, and is spoiled by the inclusion of the quadrupole parameter reported in the literature.
Note that this is not a small effect, but on the contrary, it is a very visible change in the fourth-order cumulant, that does not reach zero even in the most central collisions, as clearly observed in STAR data. 
This occurs for both $p=0$ and $p=1$.
I would like to stress, once more, that the analysis reported here is exhaustive, in the sense that one can not easily argue that negative values of $v_2\{4\}^4$ can be obtained with $\beta_2\sim-0.12$ in a different model setup.
As an example, a slightly different class of models are those models that can be fitted by a negative $p$ in \trento{}, for instance, the Monte Carlo KLN model~\cite{Drescher:2006ca}, or the Monte Carlo rcBK~\cite{ALbacete:2010ad} model.
But there the problem would remain, as these models present both a larger eccentricity~\cite{Moreland:2014oya} and a larger $\varepsilon_2\{4\}$~\cite{Giacalone:2017uqx} than the $p=0$ case.

The bottom line of this section is that the most notable effects of nuclear deformation are visible in $v_2\{4\}$, rather than in $v_2\{2\}$.
The typical statement is that nuclear deformation yields larger fluctuations of $v_2$, and thus a larger $v_2\{2\}$.
My analysis clearly indicates that experimental data are now precise enough to show distinct signatures of nuclear deformation in the details of the flow fluctuations: $v_2\{4\}^4$ in positive U+U collisions, and negative in Au+Au collisions.
Note that LHC data seems to point at the same phenomenon.
Indeed, preliminary ATLAS data show that $v_2\{4\}$ in Xe+Xe collisions is positive, all the way down to the most central events~\cite{ATLAS:2018iom}.
This is in contrast with Pb+Pb data, where a change of sign of the fourth-order cumulant of elliptic flow is currently observed~\cite{ATLAS:2017zcm}, and $^{208}$Pb nuclei are perfectly spherical\footnote{I should stress that the precise mechanism that leads to the change of sign of this cumulant in Pb+Pb collisions is not understood yet. It seems to be a generic feature of systems presenting small eccentricity driven by impact parameter fluctuations~\cite{Zhou:2018fxx}.}.

How is it, then, that $v_2\{4\}$ is so sensitive to the deformed nuclear shapes?


\section{\label{sec:4} Non-Gaussian fluctuations}

Flow fluctuations are to a good approximation Gaussian, in the sense that the elliptic flow vector $(v_x,v_y)$ has a two-dimensional Gaussian distribution in a given centrality bin.
The breakdown of Gaussian behavior leaves observable consequences in the higher-order cumulants of elliptic flow.
I refer to Refs.~\cite{Giacalone:2016eyu,Abbasi:2017ajp,Bhalerao:2018anl} for exhaustive calculations explaining how the non-Gaussian nature of fluctuations manifests in higher-order cumulants, and to the detailed analyses of Refs.~\cite{Aad:2014vba,Sirunyan:2017fts,Acharya:2018lmh,Adare:2018zkb} for experimental confirmations of such results.

In this section, I show that nuclear deformation can yield pronounced non-Gaussianity in the event-by-event distribution of elliptic flow, and I discuss the observable consequences of such an effect.

\subsection{Deformation as a source of non-Gaussianity}

Let me start by looking at the sensitivity of the cumulants $\varepsilon_2\{2\}^2$ and $\varepsilon_2\{4\}^4$ under variations of the parameter $\beta_2$, using the \trento{} model with $p=0$ and $k=0.5$.
I simulate U+U collisions at zero impact parameter, $b=0$, and I systematically vary the quadrupole deformation of the nuclei.
Note that colliding at $b=0$ implies that $\varepsilon_2\{2\}^2$ is equal to the variance of the distribution of eccentricity, whereas $\varepsilon_2\{4\}^4$ measures the kurtosis, i.e., whether the distribution has heavier or lighter tails than a Gaussian.
In particular, at $b=0$ one expects $\varepsilon_2\{4\}^4=0$ if the distribution of the eccentricity vector is a two-dimensional Gaussian.

\begin{figure}[t!]
\centering
\includegraphics[width=.8\linewidth]{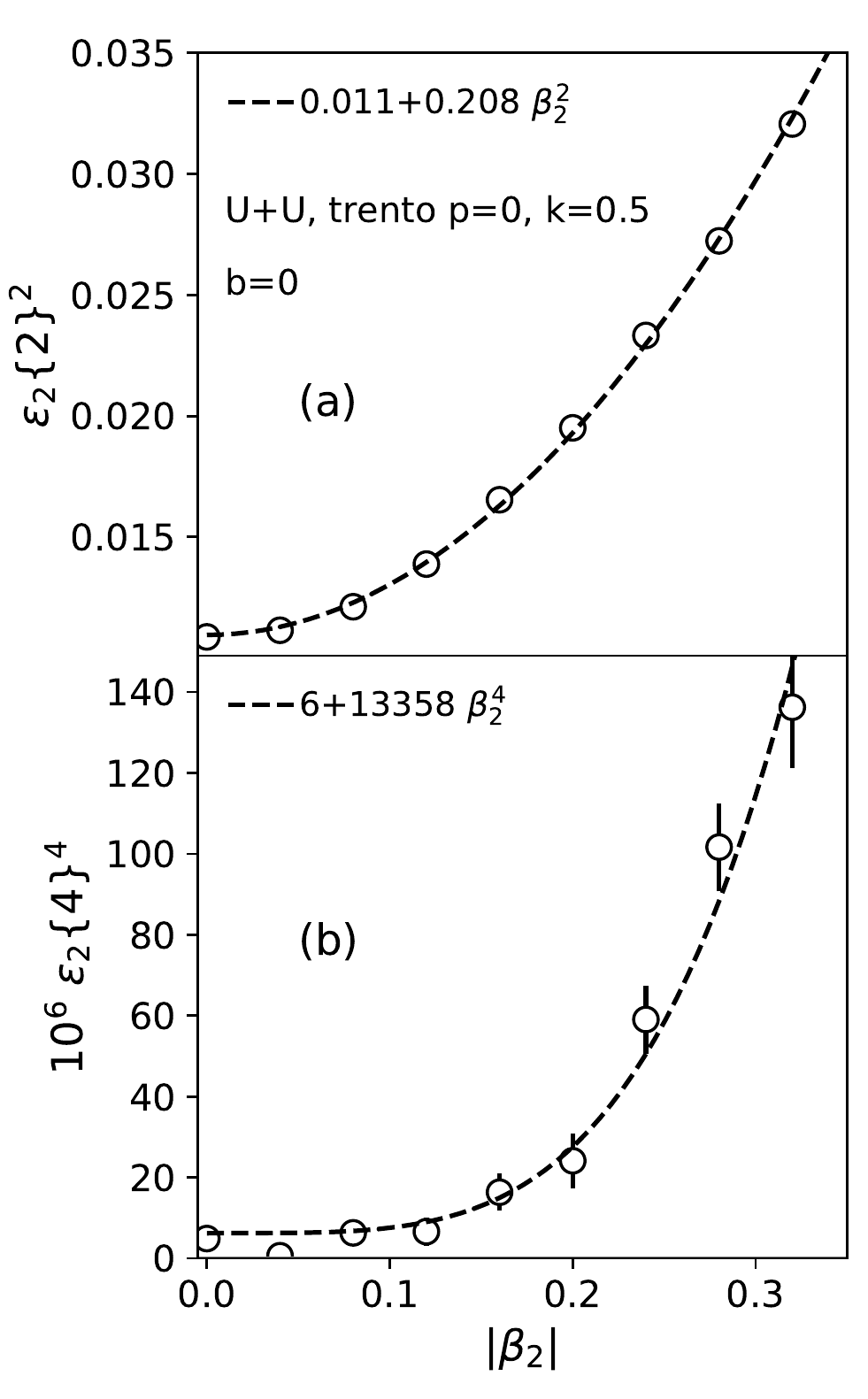}
\caption{Cumulants of eccentricity fluctuations in U+U collisions at zero impact parameter. The cumulants are plotted as function of the quadrupole coefficient, $\beta_2$. I simulate collisions using the \trento{} model with $p=0$ and $k=0.5$. Panel (a): Variance of eccentricity fluctuations. Panel (b): Kurtosis of eccentricity fluctuations. Lines are polynomial fits.}
\label{fig:3}
\end{figure}

In Fig.~\ref{fig:3}(a) I display $\varepsilon_2\{2\}^2$ as function of $\beta_2$ for collisions at zero impact parameter.
By the symmetry properties of the spherical harmonic multiplying $\beta_2$ in Eq.~(\ref{eq:ws}), any effect that is linear in $\beta_2$ should cancel when averages over events are taken.
Therefore, the value of $\bra \varepsilon_2^2 \ket$ is expected to grow with the square of $\beta_2$.
To check this, I perform a parabolic fit of $\varepsilon_2\{2\}^2$, reported as a dashed line in Fig.~\ref{fig:3}(a).
The fit is of excellent quality, and it shows how the variance of the distribution varies with the quadrupole coefficient.

Figure~\ref{fig:3}(b) shows instead $\varepsilon_2\{4\}^4$ as function of $\beta_2$.
I note a great enhancement of this quantity with increasing $\beta_2$.
This implies that nuclear deformation does not simply make the distribution of eccentricity broader, but also less Gaussian, as it makes the kurtosis grow by essentially orders of magnitude. 
The dashed line is a quartic fit, which again confirms the symmetry argument.

The previous result is very intuitive.
The kurtosis enters in $\varepsilon_2\{4\}^4$ with a negative sign~\cite{Bhalerao:2018anl}, which means that nuclear deformation makes the kurtosis of $\varepsilon_2$ fluctuations less and less negative.
Negative kurtosis for an azimuthally symmetric eccentricity distribution can be simply understood as a consequence of the fact that the eccentricity is bounded by unity~\cite{Yan:2013laa}.
Nuclear deformation, then, causes eccentricity fluctuations to get closer to their bound, and thus to present more negative kurtosis.

I conclude that nuclear deformation yields a slight increase in the variance of elliptic flow fluctuations, and a dramatic increase in their kurtosis.
This explains intuitively why $v_2\{4\}^4$ is much larger in U+U collisions than in Au+Au collision, provided, as experimental data seem to suggest, that $^{197}$Au nuclei are much more spherical than $^{238}$U nuclei.

In the following, I argue that this phenomenon has robust, observable consequences that go beyond the simple positive value of $v_2\{4\}$ in central collisions.

\subsection{Higher-order cumulants}

The bottom line of the previous discussion is that nuclear deformation yields stronger non-Gaussianity in the distribution of $v_2$.
Here I discuss the implication of this phenomenon for the splitting between higher-order cumulants of elliptic flow.

The third cumulant of elliptic flow is of order six, and is equal to
\begin{equation}
v_2\{6\} = \sqrt[6]{\frac{1}{4}\biggl ( \bra v_2^6 \ket - 9 \bra v_2^2 \ket \bra v_2^4 \ket + 12 \bra v_2^2 \ket^3 \biggr)}.
\end{equation}
Again, in the regime of linear hydrodynamic response, one can write
\begin{equation}
    v_2\{6\}=\kappa\varepsilon_2\{6\},
\end{equation}
that, using Eq.~(\ref{eq:cums}), leads to~\cite{Ma:2016hkg,Giacalone:2017uqx},
\begin{equation}
\frac{v_2\{6\}}{v_2\{4\}} = \frac{\varepsilon_2\{6\}}{\varepsilon_2\{4\}}.
\end{equation}
For collisions of spherical nuclei, e.g. $^{208}$Pb nuclei, the ratio $v_2\{6\}/v_2\{4\}$ is very close to unity ($\sim0.99$) in noncentral collisions \cite{Aad:2014vba,Sirunyan:2017fts}.
This is due to the fact that the probability distribution of elliptic flow, and consequently of the initial eccentricity, is well approximated by a two-dimensional Gaussian.
\begin{figure*}[t!]
\centering
\includegraphics[width=\linewidth]{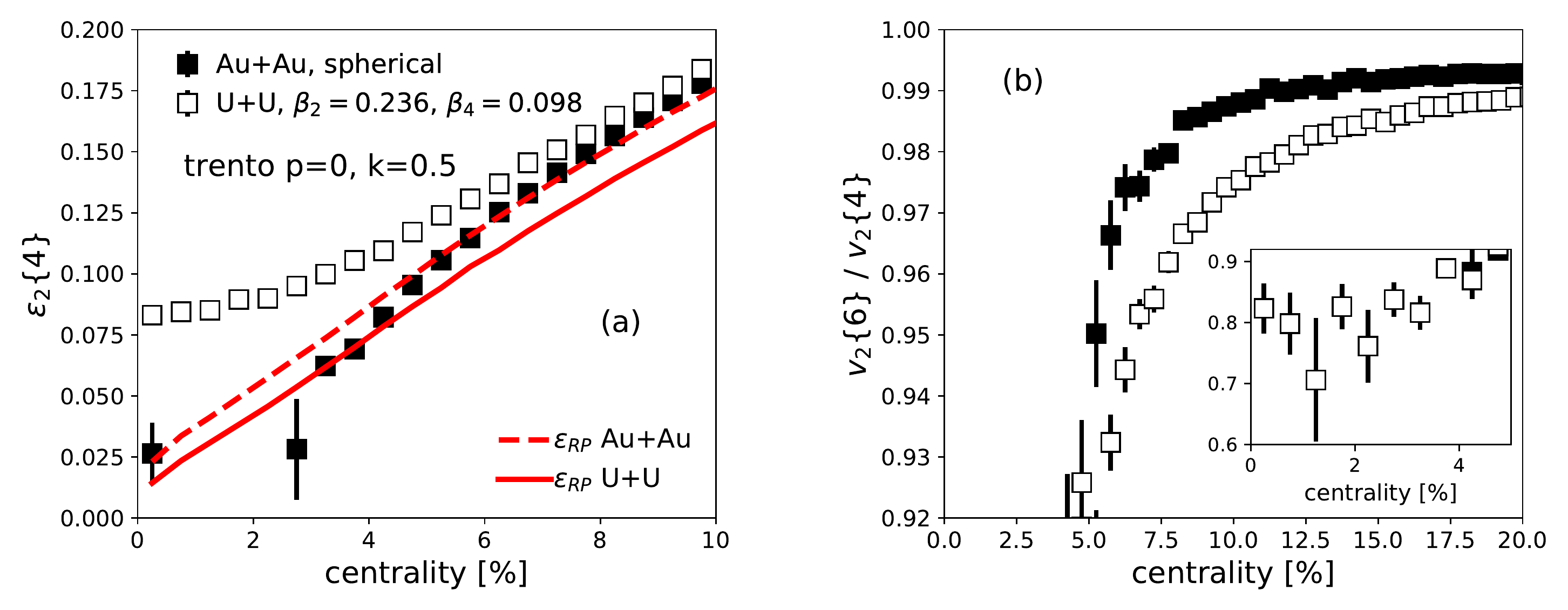}
\caption{Panel (a): $\varepsilon_2\{4\}$ in central Au+Au (full symbols) and U+U (empty symbols) collisions in the \trento{} model with $p=0$. Lines indicate the mean eccentricity in the reaction plane, $\varepsilon_{\rm RP}$, in both Au+Au (dashed line), and U+U (solid line) collisions. Panel (b): Model predictions for the ratio $v_2\{6\}/v_2\{4\}$ as function of centrality percentile, in central Au+Au (empty symbols) and U+U (full symbols) collisions. The inset is a zoom below 5\% centrality.}
\label{fig:4}
\end{figure*}
Gaussian fluctuations imply that all higher-order cumulants of $\varepsilon_2$ are equal to the mean value of the eccentricity projected along the reaction plane~\cite{Voloshin:2007pc}, which I dub, in a standard notation, $\varepsilon_{\rm RP}$\footnote{Following Eq.~(\ref{eq:ecc2}), the mean eccentricity along the direction of the reaction plane, which is customarily taken as the $x$ axis, is given by 
\begin{equation}
\varepsilon_{\rm RP} =\biggl\bra \frac{ \int r^2 \cos { 2 \phi }  ~S(r,\phi) r dr d\phi }{\int r^2 S(r,\phi) r dr d\phi}\biggr\ket ,
\end{equation}
where the average is over events.
},
\begin{equation}
\label{eq:degeneracy}
\varepsilon_2\{4\}\approx \varepsilon_2\{6\} \approx \varepsilon_{\rm RP}.
\end{equation}
This is the famous degeneracy of cumulants of the eccentricity, and of elliptic flow.
Departure from Gaussian behavior is expected starting from semi-central collisions, due to the negative skewness of $\varepsilon_2$ fluctuations in the reaction plane \cite{Giacalone:2016eyu}, as recently confirmed by LHC data in Pb+Pb collisions \cite{Sirunyan:2017fts,Acharya:2018lmh}.

I test the validity of Eq.~(\ref{eq:degeneracy}) in the \trento{} model with $p=0$.
I show in Fig.~\ref{fig:4}(a) the values of $\varepsilon_2\{4\}$ and $\varepsilon_{\rm RP}$ in central Au+Au and U+U collisions.
I choose Au+Au collisions implementing spherical nuclei, in order to highlight how results change between collisions of spherical nuclei, and collisions of non-spherical nuclei.
For Au+Au collisions, I see, as expected, that $\varepsilon_2\{4\}$ is essentially equal to $\varepsilon_{\rm RP}$ already at 5\% centrality.
This is the onset of Gaussian fluctuations.
Moving on to U+U collisions, I observe that the splitting between $\varepsilon_2\{4\}$ and $\varepsilon_{\rm RP}$ is much larger than in Au+Au collisions.
This nicely illustrates how this cumulant, in the most central events, becomes fully dominated by the enhancement of the kurtosis of the distribution, which is due to the fluctuations of the orientation of the colliding nuclei.
In these collisions, then, I do not observe any onset of Gaussian fluctuations, at least up to 10\% centrality.
I conclude that nuclear deformation breaks the degeneracy of cumulants, Eq.~(\ref{eq:degeneracy}), in central events.

Therefore, the robust, model-independent prediction I can easily draw is the following: In collisions of deformed nuclei, I expect a large splitting between $v_2\{4\}$ and $v_2\{6\}$ between $\sim$5\% centrality and semi-central collisions.
In Fig.~\ref{fig:4}(b) I show predictions for the ratio $v_2\{6\}/v_2\{4\}$, up to 20\% centrality.
Note that for U+U collisions, Gaussian fluctuations, i.e., $v_2\{6\}/v_2\{4\}\sim0.99$ are not observed below 15\% centrality.
This confirms my expectation: The ratio is significantly lower in U+U collisions than in Au+Au collisions, and the effect is very visible.
Experimental verification of this feature would provide additional confirmation of the great robustness of the hydrodynamic modeling.

Note that, from the inset in Fig.~\ref{fig:4}(b), I also predict that $v_2\{6\}^6$, much as $v_2\{4\}^4$, is positive in U+U collisions all the way to the ultra-central events.


\section{\label{sec:5} conclusions}
I have shown that the largest effects of nuclear deformation are hidden into the tails of the distribution of elliptic flow in central collisions.
Nuclear deformation yields broader distribution of elliptic flow, and, for central collisions, it engenders the mechanism of enhanced negative kurtosis presented in Sec.~\ref{sec:4}, which allows me to predict and explain the behavior of the higher-order cumulants of $v_2$.
I have, thus, explained in very simple terms the striking observation that $v_2\{4\}^4$ is much larger in U+U collisions than in Au+Au collisions, and predicted that the splitting between $v_2\{6\}$ and $v_2\{4\}$ is larger in collisions of deformed nuclei.

My results imply that there is no room in the experimental data for a $\beta_2$ of order 0.12 in the $^{197}$Au nuclei, as it would lead to too large values of the cumulant $v_2\{4\}^4$ in a model-independent way.
Clearly, one would rather trust the numbers provided in established nuclear physics literature, instead of the crude models of initial conditions for heavy-ion collisions.
But, arguably, this would just amount to rejecting evidence.
Signatures of nuclear deformation are observed in elliptic flow data in U+U collisions, and, more remarkably, in collisions of $^{129}$Xe nuclei, that present a $\beta_2$ of order $16\%$, which is very close to the reported $12\%$ of $^{197}$Au nuclei.
As I have shown in this paper, in the hydrodynamic framework one can understand all these observations as simply due to generic and model-independent features of the fluctuations of the initial geometry.
Therefore, any such effects should show up as well in Au+Au data.

This is an interesting puzzle, that, I think, nicely underlines the close link between flow fluctuations in central heavy-ion collisions, whose origin is purely geometric, and the actual shape and structure of the colliding nuclei.
This may lead to interesting developments in the future, aimed at matching these two very different areas of nuclear physics.
It would be useful, for instance, to collide nuclei whose structure and deformation is precisely determined experimentally.


\section{acknowledgements}
I thank Jean-Yves Ollitrault for useful input and illuminating discussions.
I thank Bj\"orn Schenke and Prithwish Tribedy for providing me with the results of IP-Glasma simulations, and for useful comments on the manuscript. 
I also thank Jamie Nagle for useful comments on the first draft of this paper.

\end{document}